\documentclass[useAMS,usenatbib]{mn2e}
\usepackage{graphicx}

\def\BE{\begin{equation}}
\def\EE{\end{equation}}
\def\BEA{\begin{eqnarray}}
\def\EEA{\end{eqnarray}}

\def\epsilon{\varepsilon}

\def\gtrsim{\ga}

\def\apj{{ApJ}}

\def\pasj{{PASJ}}
\def\nat{{Nature}}

\newcommand{\order}{{\cal O}}

\newcommand{\gtsima}{$\; \buildrel > \over \sim \;$}
\newcommand{\ltsima}{$\; \buildrel < \over \sim \;$}
\newcommand{\simgt}{\lower.5ex\hbox{\gtsima}}
\newcommand{\simlt}{\lower.5ex\hbox{\ltsima}}

\title[Multi-dimensional Effects]
{Multi-dimensional Treatment of Photon
Emission from Accretion
Disks 
around Black Holes}
\author[K. Kohri et al.]
{Kazunori Kohri,$^{1,2}$\thanks
{E-mail: k.kohri@lancaster.ac.uk}
Ken Ohsuga$^{3}$
and Ramesh Narayan,$^{1}$\\
$^{1}$Harvard-Smithsonian Center for Astrophysics,
MS-51, 60 Garden Street, Cambridge, MA 02138, USA\\
$^{2}$Physics Department, Lancaster University, Lancaster,LA1 4YB,UK, \\
$^{3}$Department of Physics, Rikkyo University, Toshimaku,  Tokyo
171-8501, Japan\\
}
\begin{document}

\date{
Accepted YYYY MMM DD.  
Received YYYY MMM DD; 
in original form 2006 Nov. 29 
}

\pagerange{\pageref{firstpage}--\pageref{lastpage}} \pubyear{2002}

\maketitle

\label{firstpage}

\begin{abstract}
    We consider photon emission from a supercritical accretion disk in
    which photons in the inner regions are trapped and advected in
    towards the center.  Such disks are believed to be present in
    failed supernovae and gamma-ray bursts, and possibly also in
    ultra-luminous X-ray sources.  We show that the luminosity from a
    supercritical accretion disk is proportional to the logarithm of
    the mass accretion rate when the vertical profile of the matter
    density is exponential.  We obtain analytical results also for
    other density profiles, but the dependence of the luminosity on
    the accretion rate is no longer logarithmic.
\end{abstract}

\begin{keywords}
accretion, accretion disks --- black hole physics ---
hydrodynamics --- radiative transfer
\end{keywords}

%%%%%%%%%%%%%%%%%%%%%%%%%%%%%%%%%%%%%%%%%%%%%%%%%%%%%%%%%%%%%%%%%%%%%%%
\section{Introduction}
%%%%%%%%%%%%%%%%%%%%%%%%%%%%%%%%%%%%%%%%%%%%%%%%%%%%%%%%%%%%%%%%%%%%%%%
%\label{sec:Intro}

There is wide consensus that active galactic nuclei (AGNs), gamma-ray
bursts (GRBs), X-ray binaries, etc., are powered by accretion flows on
to compact relativistic objects, most often a black hole (BH).  A
well-known model of such accretion flows is the standard-disk model
proposed by Shakura \& Sunyaev (1973), which has been used to explain
a variety of observations of AGNs and X-ray binaries.

Recently, it has been recognized that observations of many objects
cannot be explained by the standard-disk model.  For instance, this
model cannot produce the very high temperature ($T > 10^{10}$ K) and
broad-band spectrum (extending from $10^9$ to $>10^{18}$ Hz) observed
in the Galactic Center source Sgr A$^*$ and in X-ray binaries in the
hard state.  This has led to the idea of an advection-dominated
accretion flow (ADAF, Narayan \& Yi 1995, Abramowicz et
al. 1995~\footnote{We note that Ichimaru (1977) proposed similar ideas
20 years earlier.}), which is also called a radiatively inefficient
accretion flow (RIAF).  The RIAF model explains the low luminosity,
high temperature and optically-thin emission of these
systems.~\footnote{For recent developments on the RIAF model of
Sgr.~A$^{*}$, see Manmoto, Kusunose \& Mineshige. (1997), Yuan, Cui \&
Narayan (2005) and references therein.}

The standard disk cannot also be applied to very high-luminosity
accretion disks in which the mass-accretion rate $\dot{M}$ exceeds the
critical mass-accretion rate $\dot{M}_{\rm crit}$ $(\equiv L_{\rm
E}/c^{2}$, where $L_{\rm E}$ is the Eddington luminosity).  In this
regime, the disk becomes optically very thick and as a result photons
are partially trapped inside the accreting gas.  This is the regime of
interest to us.

Photon-trapping is important at any radius where the accretion time
scale is longer than the photon diffusion time scale from the disk
midplane to the surface (Begelman 1978).  The photons produced in this
region of the disk are advected into the central BH and are unable to
escape from the flow.  Begelman considered only spherically symmetric
accretion, but it is recognized that photon-trapping is important even
for systems with a disk geometry.

Photon-trapping has been included approximately in the so-called
``slim disk model'' proposed by Abramowicz, Igumenshchev \& Lasota
(1998), and in recent calculations by Szuszkiewicz et al. (1996) and
Watarai et al. (2000), as well as in numerical simulations by Honma et
al. (1991), Szuszkiewicz \& Miller (1997), and Watarai \& Mineshige
(2003).  However, these studies do not model the effect fully since
they consider vertically integrated quantities in the disk and reduce
the problem to a one-dimensional model.  Full two-dimensional
radiation-hydrodynamical simulations (2D-RHD) were done for the first
time by one of the current authors in Ohsuga et
al. (2005)~\footnote{Eggum et al. (1988) performed the first 2D
numerical simulations of supercritical accretion disks, and their work
was later improved by Okuda (2002).  However, the computation time in
these studies was short and the numerical models did not reach steady
state. Ohsuga et al. (2005) were the first to successfully compute
models approaching steady state.}.

A realistic model of supercritical accretion must include the
following two effects: i) Even at disk radii inside the trapping
radius, photons that are emitted near the surface of the disk can
escape, though most of the photons emitted deeper inside are advected
into the BH.  ii) The vertical profiles of various quantities in the
disk such as density, optical depth, etc., play a critical role in
determining what fraction of the photons escape, and thereby influence
the total luminosity of the disk.

The first effect was pointed out and studied by Ohsuga et al. (2002)
using simple analytical models and numerical simulations. In this
paper, we include the second effect and study the combined influence
of both effects for a broad class of accretion-disk models.  We
develop an analytical model and show that it agrees well with the
numerical results of Ohsuga et al. (2005).

The results obtained here may change the current understanding of
models such as the slim-disk and convection-dominated accretion flow
model (CDAF, Stone, Pringle \& Begelman 1999; Narayan, Igumenshchev \&
Abramowicz 2000; Quataert \& Gruzinov 2000; Igumenshchev, Abramowicz
\& Narayan 2000), as well as the neutrino-dominated accretion flow
(NDAF) model (Popham, Woosley \& Fryer. 1999; Narayan, Piran \& Kumar
2001; Kohri \& Mineshige 2002; Di Matteo, Perna \& Narayan 2002;
Kohri, Narayan \& Piran 2005 Gu, Liu \& Lu, 2006) which is believed to
play an important role in the central engine of GRBs (Narayan,
Paczynski \& Piran 1992).~\footnote{The relativistic
magnetohydrodynamic (MHD) jet model is another attractive model of
energetic accretion sources (e.g., McKinney 2005; 2006 and references
therein).  This model is less likely to be affected by our work.}
%That is because the most dominant cooling process among the
%radiative, the advective, the convective and the neutrino coolings is
%apparently affected by the multi-dimensional effects.

In the following we use $R$ to represent the cylindrical radius of a
point with respect to the central mass and $z$ to represent its
vertical distance from the equatorial plane of the disk. We also
assume that the temperature $T$ in the disk is lower than the electron
mass ($kT < m_{e}c^2$), so that the electrons are nonrelativistic.

%%%%%%%%%%%%%%%%%%%%%%%%%%%%%%%%%%%%%%%%%%%%%%%%%%%%%%%%%%%%%%%%%%%%%%%
\section{Standard one-dimensional approximations}
%%%%%%%%%%%%%%%%%%%%%%%%%%%%%%%%%%%%%%%%%%%%%%%%%%%%%%%%%%%%%%%%%%%%%%%

In this section we introduce the standard one-dimensional (1D)
treatment of accretion disks, following the approach described in
Ohsuga et al (2002). In this approach, we integrate along the $z$-axis
and omit the $z$-dependences of various astrophysical quantities. For
instance, we write the optical depth from the mid-plane of the disk
($z=0$) to the surface ($z=z_{\rm max}$) as
\begin{eqnarray}
    \label{eq:tau_1_1}
    \tau &=& \int_{0}^{z_{\rm max}} dz~ \sigma_{\gamma e} n_{e} \\ 
    \label{eq:tau_1_2}
         &\simeq& \sigma_{\rm T} n_{e} H,
\end{eqnarray}
where $\sigma_{\gamma e}$ is the scattering cross section of a photon
off an electron. Although $\sigma_{\gamma e}$ depends in general on
the photon energy $E_{\gamma}$, we limit ourselves to Thomson
scattering, for which the cross-section is independent of energy.
This is reasonable for the low energy photons $E_{\gamma} \ll
\order({\rm MeV})$ of interest to us.  $H$ is the disk half-thickness
and $n_{e}$ is the electron number density in the disk mid-plane.
%Here we assume
%that the ratio $H/R$, where $R$ is  the radial coordinate does not
%depend strongly on $R$, i.e., $H/R$ is either small or constant.  This
%approximation is  reasonable because we mainly consider   thin disks
%(or a mixture of a standard radiatively cooled  thin disk and an
%advection dominated accretion disk).  

The diffusion time scale for a photon to escape from the disk mid-plane
is given by
\begin{eqnarray}
    \label{eq:tdiff_1}
    t_{\rm diff} \approx N_{\rm scatt} \ \lambda /c,
\end{eqnarray}
where $N_{\rm scatt}$ is the number of scatterings, $\lambda =
1/(\sigma_{\rm T} n_{e})$ is the mean free path, and $c$ is the speed
of light. Since $N_{\rm scatt}$ is given by
\begin{eqnarray}
    \label{eq:Nscatt_1}
    N_{\rm scatt} \approx 3~\tau^{2},
\end{eqnarray}
(assuming a random walk in three-dimensions, see
Appendix~\ref{sec:deriv_tdiff}), we write
\begin{eqnarray}
    \label{eq:t_diff_1}
    t_{\rm diff} \approx 3 \tau H/ c.
\end{eqnarray}

The accretion time scale is
\begin{eqnarray}
    \label{eq:tacc}
    t_{\rm acc} = - \frac{R}{ v_{r}},
\end{eqnarray}
where the radial velocity $v_{r}$ is related to the mass accretion
rate $\dot{M}$ and the density $\rho$ by
\begin{eqnarray}
    \label{eq:vr}
    v_{r} &=& - \displaystyle{\frac{\dot{M}} 
    {4 \pi R \int_{0}^{H}\rho dz }}  \\
      &=& - \displaystyle{\frac{\dot{M}} 
    {4 \pi R  \rho H}}.
\end{eqnarray}
Thus
\begin{eqnarray}
    \label{eq:tacc1}
    t_{\rm acc} \approx \frac{2\tau}{\dot{m}}\frac{R^{2}}{c  \  R_{g} },
\end{eqnarray}
where $R_{g} = 2 G M/c^{2}$ is the Schwarzschild radius, and we have
set $\rho \approx m_{p} n_{e}$ and introduced a dimensionless mass
accretion rate $\dot{m}$,
\begin{eqnarray}
    \label{eq:smlmdot}
    \dot{m} = \dot{M} / \dot{M}_{\rm crit}.
\end{eqnarray}
The critical mass accretion rate $\dot{M}_{\rm crit}$ is defined in
terms of the Eddington luminosity $L_{\rm E}$ as follows,
$\dot{M}_{\rm crit} = L_{\rm E}/c^{2}$, where $L_{\rm E} = {4 \pi c G
m_{p}}M /{\sigma_{\rm T}} \simeq 1.3 \times 10^{38} {\rm erg}~{\rm
sec}^{-1} (M/M_{\odot})$ for a black hole of mass $M$.

Photons can escape from the disk only if  the condition $t_{\rm diff}
< t_{\rm acc}$ is satisfied. Therefore, the disk  radiates freely only
from radii larger than a certain limit,
\begin{eqnarray}
    \label{eq:largeR}
    R >  R_{\rm trap},
\end{eqnarray}
where the trapping radius $R_{\rm trap}$ is given by
\begin{eqnarray}
    \label{eq:Rtrap}
    R_{\rm trap} =   \frac32 h_{\rm in} \dot{m} R_{g}.
\end{eqnarray}
and
\begin{eqnarray}
    \label{eq:small_h}
    h_{\rm in} \left. \equiv \frac{H}{R} \right|_{R \le R_{\rm trap}},
\end{eqnarray}
assuming that, when $R<R_{\rm trap}$ and radiation is trapped, the
disk is geometrically thick with $h_{\rm in}\sim1$ and that $h$ here
is independent of $R$.  As we show in
Appendix~\ref{sec:disk-half-thickness}, $h$ is indeed constant
($\approx$ 0.5) for $R \le R_{\rm trap}$ and decreases $\propto 1/R$
for $R_{\rm trap} < R$. 
%In the outer region, therefore, the diffusion
%time scale which is a increasing function of $h$ is actually shorter
%than what one estimates assuming constant $h$.  
This validates
Eq.~(\ref{eq:Rtrap}).  Note that $R_{\rm trap}$ is proportional to
$\dot{m}$.

The energy released in the disk per unit time by viscous dissipation
is given by
\begin{eqnarray}
    \label{eq:lum_1}
    L =  2 \int_{R_{\rm in}}^{R_{\rm out}} 2 \pi R Q_{\rm vis} (R) dR,
\end{eqnarray}
where the viscous heating rate is 
\begin{eqnarray}
    \label{eq:Qvis}
    Q_{\rm vis } (R) = \frac{3}{8\pi} \frac{GM}{R^{3}} \dot{M} \left[1
    - \left( \frac{R_{\rm in}}{R} \right)^{1/2} \right].
\end{eqnarray}
Here $R_{\rm in}$ and $R_{\rm out}$ are the radii of the inner and
outer edge of the disk.  However, not all the released energy is
radiated because of photon-trapping.  We discuss two cases below,
depending on whether $R_{\rm trap}$ is larger or smaller than $R_{\rm
in}$. For readers' convenience we introduce a dimensionless inner
radius normalized by the Schwarzschild radius.
\begin{eqnarray}
    \label{eq:smallrin}
    r_{\rm in} \equiv R_{\rm in} /R_{\rm g}.
\end{eqnarray}
%%

%%%%%%%%%%%%%%%%%%%%%%%%%%%%%%%%%%%%%%%%%%%%%%%%%%%%%%%%%%%%%%%%%%%%%%%
\subsection{$R_{\rm trap} < R_{\rm in}$}
%%%%%%%%%%%%%%%%%%%%%%%%%%%%%%%%%%%%%%%%%%%%%%%%%%%%%%%%%%%%%%%%%%%%%%%
\label{subsec:notrap}

When $R_{\rm trap} < R_{\rm in}$, there is no trapped region and we
estimate the total luminosity to be
\begin{eqnarray}
    \label{eq:lum_2_1}
    L_{{\rm 1D, tot},R_{\rm trap} < R_{\rm in}}
    &=& 
    2 \int_{R_{\rm in}}^{\infty} 2 \pi R Q_{\rm vis} (R) 
    dR \nonumber \\
    &=& \frac1{6 h_{\rm in}} L_{\rm E}  \frac{R_{\rm trap}}{R_{\rm in}}
\nonumber \\
    &=& \frac{L_{\rm E}}{4 r_{\rm in}} \dot{m},
\end{eqnarray}
which is linearly proportional to $\dot{m}$.  This is a well known
result for the standard thin accretion disk model in Newtonian
gravity.

%%%%%%%%%%%%%%%%%%%%%%%%%%%%%%%%%%%%%%%%%%%%%%%%%%%%%%%%%%%%%%%%%%%%%%%
\subsection{$R_{\rm in} < R_{\rm trap}$}
%%%%%%%%%%%%%%%%%%%%%%%%%%%%%%%%%%%%%%%%%%%%%%%%%%%%%%%%%%%%%%%%%%%%%%%
\label{subsec:sometrap}

When $R_{\rm in} < R_{\rm trap}$, only the region of the disk outside
$R_{\rm trap}$ radiates freely and the luminosity from this region of
the disk is
\begin{eqnarray}
    \label{eq:lum_2_2_1}
   L_{{\rm 1D}, R_{\rm trap}<R} 
  &=& 2 \int_{R_{\rm trap}}^{\infty} 
  2 \pi R Q_{\rm vis} (R) dR 
\nonumber  \\
      &=& \frac1{2h_{\rm in}} L_{\rm E} 
      \left[1 - \frac23 
        \left(\frac{R_{\rm trap}}{R_{\rm in}} 
      \right)^{-1/2}\right]
\nonumber  \\
      &=& \frac1{2h_{\rm in}} L_{\rm E} 
      \left[1 - 
        \sqrt{\frac{8r_{\rm in}}{27h_{\rm in}}}~\dot{m}^{-1/2}\right].
\end{eqnarray}
Within the one-dimensional approximation being discussed in this
section, it is unclear how much luminosity is emitted from the region
of the disk $R_{\rm in} < R < R_{\rm trap}$.  One extreme assumption
is that there is no luminosity at all from this region, as in Ohsuga
et al. 2002.  Alternatively, one might assume that the disk flux here
is limited to the local Eddington flux $F \approx F_{\rm E}(R) =
L_{\rm E}/(4\pi R^{2}) = G M m_{p} c / (R^{2}\sigma_{\rm T} )$.  The
luminosity from $R_{\rm in} < R < R_{\rm trap}$ of the disk is then
given by
\begin{eqnarray}
    \label{eq:lum_2_2_2}
    L_{{\rm 1D}, R<R_{\rm trap}}
  &=& 2 \int_{R_{\rm in}}^{R_{\rm trap}} 
  2 \pi R F_{\rm Edd}(R)  dR 
 \nonumber\\
      &=& L_{\rm E} \ln\left(\frac{R_{\rm trap}}{R_{\rm in}}
      \right)
 \nonumber\\
      &=& L_{\rm E} \ln\left(\frac{3 h_{\rm in}}{2 r_{\rm in}}\dot{m}
      \right).
\end{eqnarray}

If we include the term given in Eq.~(\ref{eq:lum_2_2_2}), keeping in
mind that it is uncertain, then the ``total luminosity'' becomes
\begin{eqnarray}
    \label{eq:lum_2_2}
   \lefteqn{L_{\rm 1D,tot,R_{\rm in} < R_{\rm trap}}\nonumber } \\ 
    &=& L_{{\rm 1D}, R_{\rm trap}<R} 
   + L_{{\rm 1D}, R<R_{\rm trap}} 
   \nonumber \\
      &=& \frac1{2h_{\rm in}} L_{\rm E} 
      \left[1 + 2 h_{\rm in}\ln\left(\frac{R_{\rm trap}}{R_{\rm in}}
      \right)  
      - \frac23 
        \left(\frac{R_{\rm trap} }{R_{\rm in}} 
      \right)^{-1/2}\right]
    \\ \nonumber
      &=& \frac1{2h_{\rm in}} L_{\rm E} 
      \left[1 + 2 h_{\rm in}\ln\left(\frac{3 h_{\rm in}}{2 r_{\rm in}}\dot{m}
      \right)
      - \sqrt{\frac{8r_{\rm in}}{27h_{\rm in}}}~\dot{m}^{-1/2}\right].
\end{eqnarray}
Note that the luminosity is reduced relative to the result in
Eq.~(\ref{eq:lum_2_1}) for the untrapped case, and increases only
logarithmically with $\dot{m}$.  This explains why disks with
supercritical accretion ($\dot{m}\gg1$, ``slim'' disks) are
radiatively inefficient.

%%%%%%%%%%%%%%%%%%%%%%%%%%%%%%%%%%%%%%%%%%%%%%%%%%%%%%%%%%%%%%%%%%%%%%%
\section{Two-dimensional Treatment}
%%%%%%%%%%%%%%%%%%%%%%%%%%%%%%%%%%%%%%%%%%%%%%%%%%%%%%%%%%%%%%%%%%%%%%%
\label{sec:2dim}

So far we have discussed a 1D approximation and assumed somewhat
arbitrarily that the region $R < R_{\rm trap}$ either does not radiate
at all or radiates at the local Eddington rate.  In this section we
carry out a more careful two-dimensional analysis.  We focus on the
region $R_{\rm in} \le R < R_{\rm trap}$ discussed in
Subsection~\ref{subsec:sometrap}, but now we carefully consider the
two-dimensional distribution of quantities.

In the following, we consider most of the quantities in the disk to
depend on both $R$ and $z$ (the vertical height).  However, we assume
that the thin-disk approximation is approximately valid in the sense
that we omit the $z$-dependence when considering hydrostatic balance
in the $z$-direction ($c_{S} = \Omega_{K} H $, with the sound speed
$c_{S}$ and the Keplerian angular velocity $\Omega_{K} =
\sqrt{GM/R^{3}}$) and angular momentum conservation ($\nu \int \rho dz
= \dot{M}(1-\sqrt{R_{\rm in}/R})/(3\pi)$, with the kinetic viscosity
$\nu$).  Then the $z$-dependent optical depth at a given $R$ is given
by
\begin{eqnarray}
    \label{eq:tauz}
    \tau(z) &=& \int_{z}^{z_{\rm max}} dz~ \sigma_{\gamma e} n_{e} (z) \\
    \label{eq:tauz_2}
          &\simeq& \sigma_{\rm T} \int_{z}^{z_{\rm max}} dz~ n_{e} (z),
\end{eqnarray}
where $n_{e}(z)$ is the electron number density at height $z$, and
$z_{\rm max}$ corresponds to the upper surface of the disk. We see
that the expression for $\tau$ given in Eq.~(\ref{eq:tau_1_1}) in the
1D approximation corresponds to $\tau(0)$ in Eq.~(\ref{eq:tauz}).

\begin{figure}
%\epsscale{1.0}
%\includegraphics[width=80mm]{x20_aug8.ps}
%\includegraphics[width=95mm]{x20_dec18.ps}
%\includegraphics[width=95mm]{x20_jan27.ps}
\includegraphics[width=90mm]{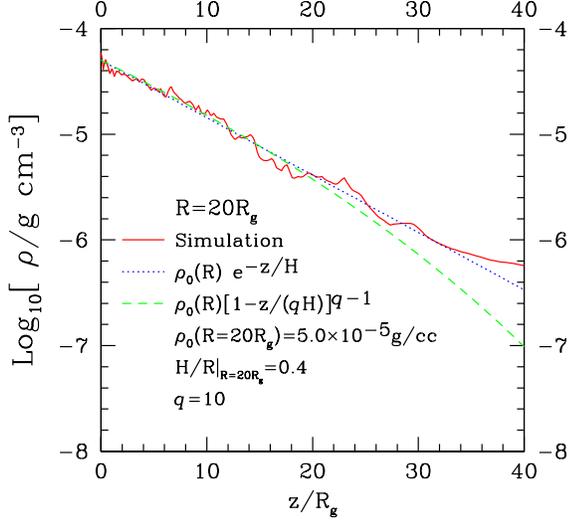}
%\plotone{x20_aug8.ps}
\vspace{.0in}
\caption{ Variation of the matter density $\rho$ with height $z$ at a
radius $R = 20 R_{\rm g}$.  The solid line represents the result from
the numerical simulation of Ohsuga et al. (2005).  The data are fit
well with either an exponential (dotted line) or a power-law (dashed)
profile, with $H/R$ = 0.4 and (for the power-law profile) $q = 10$.
The numerical results in Ohsuga et al. (2005) can be fit with the same
value of $H/R$ = 0.4 for all radii up to $\sim 50 R_{\rm g}$.}
\label{fig:rho_x}
\end{figure}

We now introduce a detailed model for the vertical electron density
profile. As seen in Fig.~\ref{fig:rho_x}, we can fit the the density
profile found in the numerical simulation in Ohsuga~et~al.~(2005)
quite well with the following power-law form,
\begin{eqnarray}
    \label{eq:power-law}
    n_{e}(z) = n_{e}(0) \left( 1- \frac{z}{q H}
    \right)^{q-1}
    (q \ge 1,~{\rm and}~0 \le z \le q H),
\end{eqnarray}
where $n_{e}(0)$ and the power-law index $q$ may depend on the radius
$R$.  In the limit as $q\to\infty$, this model takes the form of an
exponential~\footnote{If on the other hand we have a constant
$z$-independent density, as considered by Ohsuga et al. (2002), it
would correspond to $q = 1$.  The corresponding results can be
recovered by setting $q=1$ in all the expressions in the current
paper.}
\begin{eqnarray}
    \label{eq:exponential} n_{e}(z) = n_{e}(0) \exp\left(- \frac{z}{H}
    \right) \qquad (0 \le z \le \infty).
\end{eqnarray}
The optical depth from $z$ to the surface is
\begin{eqnarray}
    \label{eq:tauz_pow}
    \tau(z) = \tau(0) \left(1 - \frac{z}{q H} \right)^{q},
\end{eqnarray}
except for $q\to\infty$ when it takes an exponential form.

Because we adopt the numerical results corresponding to $\dot{m}
\gtrsim 100$ in Ohsuga~et~al.~(2005), the radii $R$ of order tens of $
R_{g}$ that we consider are smaller than $R_{\rm trap} \sim 10^{2}
R_{g}$ and so we are in the radiation-trapped regime.  Note that
Fig.~\ref{fig:rho_x} gives $H/R \approx 0.4$ for the numerical
simulation of Ohsuga et al. (2005), which is approximately consistent
with the approximation $h_{\rm in} = 0.5$ that we made in the previous
section and in Appendix~\ref{sec:disk-half-thickness}.

%%%%%%%%%%%%%%%%%%%%%%%%%%%%%%%%%%%%%%%%%%%%%%%%%%%%%%%%%%%%%%%%%%%%%%%%
%\clearpage
%%%%%%%%%%%%%%%%%%%%%%%%%%%%%%%%%%%%%%%%%%%%%%%%%%%%%%%%%%%%%%%%%%%%%%%

The photon diffusion time scale from height $z$ to the surface of the
disk at $z_{\rm max}$ is given by
\begin{eqnarray}
    \label{eq:tdiffz}
    t_{\rm diff}(z) \approx N_{\rm scatt}(z,z_{\rm max})~\lambda(z)/c,
\end{eqnarray}
where $N_{\rm scatt}(z,z_{\rm max})$ is the number of scatterings in
the 3D random walk,
\begin{eqnarray}
    \label{eq:nscatt_2}
    N_{\rm scatt}(z,z_{\rm max}) = 3~\left[\tau(z)\right]^{2},
\end{eqnarray}
and $\lambda(z)$ is the mean free path which is estimated using the
properties of the disk at the starting position $z$ of the photon
\begin{eqnarray}
    \label{eq:lambda_z}
    \lambda(z) =  \frac1{\sigma_{\rm T}~n_{e}(z)}.
\end{eqnarray}
The detailed derivation of the above relation is given in
Appendix~\ref{sec:deriv_tdiff}.  By using Eqs.~(\ref{eq:nscatt_2})
and~(\ref{eq:lambda_z}), we obtain an expression for the $z$-dependent
diffusion time scale,
\begin{eqnarray}
    \label{eq:tdiffz2}
  t_{\rm diff}(z) &=&  3 \tau(z)^{2} \lambda(z) /c \nonumber \\
   &=& 3 \tau(0) \frac{H}{c} 
   \left[ \frac{\tau(z)}{\tau(0)} \right]^{(q+1)/q} \,.
\end{eqnarray}

\begin{figure}
%\epsscale{1.0}
%\includegraphics[width=80mm]{diffusion_3.eps}
\includegraphics[width=85mm]{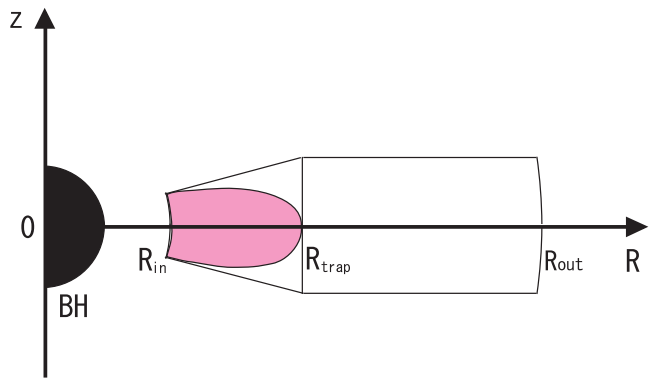}
%\plotone{diffusion_3.eps}
\vspace{.0in}
\caption{Schematic representation of a supercritical accretion disk
around a BH. The shaded region shows the photon-trapping region from
which no photons can escape. We see that, even for $R < R_{\rm trap}$,
there exist regions near the surface of the disk from which photons
can diffuse out (white region).  }
\label{fig:diffusion}
\end{figure}

By requiring the diffusion time scale $t_{\rm diff}(z)$ to be shorter
than the accretion time scale $t_{\rm acc}$ given in
Eq.~(\ref{eq:tacc1}), we find that photons can escape only from the
region  where the following condition is satisfied,
\begin{eqnarray}
    \label{eq:cond_diff_0}
    \frac{\tau(z)}{\tau(0)} < 
    \left(\frac{R}{R_{\rm trap}} \right)^{q/(q+1)},
\end{eqnarray}
The photon-trapping radius $R_{\rm trap}$ is determined by
Eq.~(\ref{eq:Rtrap}) and is proportional $\dot{m}$.  A schematic
picture indicating the region of the disk from which radiation can
escape is shown in Fig.~\ref{fig:diffusion}.  We see that trapping is
not determined by simply a critical radius $R_{\rm trap}$, but is
described in terms of a two-dimensional surface $z_{\rm trap}(R)$,
where
\begin{eqnarray}
    \label{eq:z_trap}
    \frac{\tau(z_{\rm trap})}{\tau(0)} \equiv  \left(\frac{R}{R_{\rm
trap}}\right)^{q/(q +1)}.
\end{eqnarray}
Note that $z_{\rm trap}$ is a function of $R$ in this case given by
\begin{eqnarray}
    \label{eq:ztrap_R}
    z_{\rm trap} = qH \left[ 1
          - \left(\frac{R}{R_{\rm trap}} \right)^{{1/(q+1)}}
        \right],
\end{eqnarray}
and its second-order  derivative by $R$ is always negative $d^2z_{\rm
trap}/dR^{2} <0$ for $R<R_{\rm trap}$ because
\begin{eqnarray}
    \label{eq:dztrapdR}
    \frac{dz_{\rm trap}}{dR} =
    q h_{\rm in} \left[1 - \frac{q+2}{q+1} 
    \left(\frac{R}{R_{\rm trap}} \right)^{1/(q+1)} \right],
\end{eqnarray}
and
\begin{eqnarray}
    \label{eq:d2ztrapdR2}
    \frac{d^{2}z_{\rm trap}}{dR^{2}} =
    -   \frac{q(q+2)}{q+1}  h_{\rm in}  \frac1{R_{\rm trap}}
    \left(\frac{R}{R_{\rm trap}} \right)^{-q/(q+1)},
\end{eqnarray}
for $R<R_{\rm trap}$.
The luminosity from the region $R_{\rm in} \le R \le R_{\rm trap}$ is
then given by
\begin{eqnarray}
    \label{eq:L_add_pow}
    L_{{\rm 2D}, R<R_{\rm trap}}
    = 2 \int_{R_{\rm in}}^{R_{\rm trap}}  2 \pi R Q_{\rm
    vis } (R) \frac{\tau(z_{\rm trap})}{\tau(0)}dR.
\end{eqnarray}
Here we assume that all the energy released at $z \geq z_{\rm trap}$
is radiated, while the energy released at $z < z_{\rm trap}$ is
completely trapped.  Also, because $Q_{\rm vis}$ and $\tau(z)$ are
both proportional to $\int \rho dz$, we express the vertical
distribution of viscous dissipation directly by means of the factor
$\tau(z_{\rm trap})/\tau(0)$.

%%%%%%%%%%%%%%%%%%%%%%%%%%%%%%%%%%%%%%%%%%%%%%%%%%%%%%%%%%%%%%%%%%%%%%%%
%\clearpage
%%%%%%%%%%%%%%%%%%%%%%%%%%%%%%%%%%%%%%%%%%%%%%%%%%%%%%%%%%%%%%%%%%%%%%%

Evaluating the above integral, we obtain the luminosity of the disk
from the region $R < R_{\rm trap}$:
\begin{eqnarray}
    \label{eq:L_add_pow2_1_1}
    \lefteqn{ L_{{\rm 2D},R<R_{\rm trap}}  }   \nonumber  \\ 
    \lefteqn{= 2\int_{R_{\rm in}}^{R_{\rm trap}}  2 \pi R Q_{\rm
    vis } (R)  \left(\frac{R}{R_{\rm trap}}
    \right)^{q/(q+1)} dR }  \nonumber \\
    \label{eq:L_add_pow2_2}
    \lefteqn{ = \frac1{2h_{\rm in}} L_{\rm E} 
      \left(\frac{R_{\rm trap}}{R_{\rm in}} \right)^{1/(q+1)}  }
      \nonumber \\
      \lefteqn{ \hspace{1cm} \times \int_{1}^{R_{\rm trap}/R_{\rm in}} dx 
      \frac{1}{x^{(q+2)/(q+1)}} \left(1 - \frac{1}{x^{1/2}} 
    \right) }  \nonumber \\
    \label{eq:L_add_pow2_3}
    \lefteqn{=\frac1{2h_{\rm in}} L_{\rm E} \times}  \nonumber \\
    \lefteqn{ \left[
      \frac{2(q+1)}{q+3} 
      \left(\frac{R_{\rm trap}}{R_{\rm in}} \right)^{-1/2}
      + \frac{(q+1)^{2}}{q+3} 
      \left(\frac{R_{\rm trap}}{R_{\rm in}} \right)^{1/(q+1)}
      - (q + 1)
    \right]} \nonumber
\\
    \label{eq:L_add_pow2_3_2}
    \lefteqn{=\frac1{2h_{\rm in}} L_{\rm E} \times}  \\
    \lefteqn{ \left[
      \frac{2(q+1)}{q+3} 
      \left(\frac{3 h_{\rm in} \dot{m}}{2 r_{\rm in}} \right)^{-1/2}
      + \frac{(q+1)^{2}}{q+3} 
      \left(\frac{3 h_{\rm in} \dot{m}}{2 r_{\rm in}} \right)^{1/(q+1)}
      - (q + 1)
    \right].} \nonumber
\end{eqnarray}
The dominant contribution comes from the second term, which is
proportional to $\dot{m}^{1/(q+1)}$ for $\dot{m} \gg 1$.
The luminosity from the region $R_{\rm trap} < R$ is, of course, the
same as in the 1D case (Eq.~\ref{eq:lum_2_2_1}), i.e., $L_{{\rm 2D},
R_{\rm trap}<R} = L_{{\rm 1D}, R_{\rm trap}<R}$, since there is no
photon trapping.  Therefore, we have the following expression for the
total radiated luminosity from the entire disk,
\begin{eqnarray}
    \label{eq:L_add_pow2_1}
    \lefteqn{L_{\rm 2D,tot}(q) }  \nonumber \\ 
    \lefteqn{= L_{{\rm 2D},R<R_{\rm trap}} + L_{{\rm 2D},R_{\rm
    trap}<R}  } 
    \nonumber\\
    \lefteqn{= \frac1{2h_{\rm in}} L_{\rm E}\times} \nonumber \\
    \lefteqn{ \left[ 
      \frac{ 4 q}{3(q+3)}  
        \left(\frac{R_{\rm trap}}{R_{\rm in}} \right)^{-1/2} 
         +  \frac{(q+1)^{2}}{q+3} 
      \left(\frac{R_{\rm trap}}{R_{\rm in}} \right)^{1/(q+1)}
      - q 
    \right] }  \nonumber \\ 
    \nonumber \\
    \lefteqn{= \frac1{2h_{\rm in}} L_{\rm E}\times}  \\ \nonumber
%    \label{eq:L_add_pow2_1_2}
    \lefteqn{ \left[ 
      \frac{ 4 q}{3(q+3)}  
        \left(\frac{3 h_{\rm in} \dot{m}}{2 r_{\rm in}} \right)^{-1/2} 
         +  \frac{(q+1)^{2}}{q+3} 
      \left(\frac{3 h_{\rm in} \dot{m}}{2 r_{\rm in}}\right)^{1/(q+1)}
      - q 
    \right]. } \nonumber
\end{eqnarray}

In a similar fashion, we obtain the following result for the case of
an exponential density profile ($q\to\infty$),
\begin{eqnarray}
    \label{eq:L_add_exp}
   \lefteqn{ L_{\rm 2D,tot}(q=\infty)} 
   \nonumber \\
&=& \frac1{2h_{\rm in}} L_{\rm E} 
    \left[
      \frac43 \left(\frac{R_{\rm trap}}{R_{\rm in}} \right)^{-1/2}
      + \ln\left(\frac{R_{\rm trap}}{R_{\rm in}} \right)
      -1
    \right]
    \\ \nonumber
&=& \frac1{2h_{\rm in}} L_{\rm E} 
    \left[
      \sqrt{\frac{32r_{\rm in}}{27h_{\rm in}}}~\dot{m}^{-1/2}
      + \ln\left(\frac{3 h_{\rm in}}{2 r_{\rm in}}\dot{m}
      \right)
      -1
    \right].
\end{eqnarray}
To verify that this result is consistent with the limit $q\to\infty$
of Eq.~(\ref{eq:L_add_pow2_3_2}), we make use of the following useful
relation,
\begin{eqnarray}
    \label{eq:limit_eps0}
    \lim_{\epsilon \to 0} \frac{1}{\epsilon} \left[ \frac1{2\epsilon +
    1} x^{\epsilon}-1 \right] = \ln x -2.
\end{eqnarray}
This shows that the dominant terms in Eq.~(\ref{eq:L_add_pow2_3_2}) do
not diverge as $q\to\infty$ but tend to a finite value that is
proportional to $\ln(\dot{m})$ for $\dot{m} \gg 1$.

%%%%%%%%%%%%%%%%%%%%%%%%%%%%%%%%%%%%%%%%%%%%%%%%%%%%%%%%%%%%%%%%%%%%%%%
\section{Comparison with numerical simulation}
%%%%%%%%%%%%%%%%%%%%%%%%%%%%%%%%%%%%%%%%%%%%%%%%%%%%%%%%%%%%%%%%%%%%%%%

Here we compare the analytical formulae derived above for the
luminosity $L$ of the disk with the results of multi-dimensional
numerical simulations reported by Ohsuga~et~al.~(2005).

\begin{figure}
%\epsscale{1.0}
%\plotone{lm_aug9.ps}
%\includegraphics[width=80mm]{lm_aug9.ps}
\includegraphics[width=95mm]{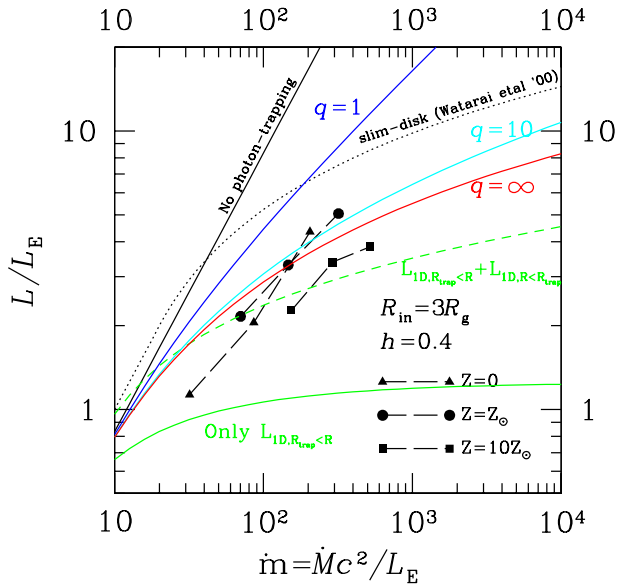}
\vspace{.0in}
\caption{Variation of the disk luminosity as a function of the
mass-accretion rate. Analytical formulae are denoted by solid
lines. From above, the lines correspond to the case with
photon-trapping ignored, power-law density profiles with $q$ = 1
(constant density), $q$ = 10, and $q\to \infty$ (exponential profile),
and a standard 1D model in which photons are assumed to be fully
trapped for $R_{\rm in} < R < R_{\rm trap}$.  The triangles, circles,
and squares connected by long-dashed lines indicate the numerical
results shown in Fig.~7 in Ohsuga~et~al.~(2005), for metallicity $Z$ =
0, $Z_{\odot}$, and 10 $Z_{\odot}$, respectively.  The dotted line
represents the case of the hydrodynamical slim-disk simulations of
Watarai et al. (2001), without explicitly including 2D effects. The
short-dashed line shows the profile of luminosity in an approximate
analytical 1D model in which we assume that the disk emits at the
local Eddington rate at radii $R_{\rm in} < R < R_{\rm trap}$.  All
results correspond to $R_{\rm in}=3 R_{\rm g}$ and $h$ = 0.4.  }
\label{fig:lm}
\end{figure}

In Fig.~\ref{fig:lm} we plot the disk luminosity in Eddington units as
a function of the dimensionless mass-accretion rate
$\dot{m}$. Analytical formulae are denoted by solid lines and
correspond, in descending order, to the following cases: no
photon-trapping (Eq.~\ref{eq:lum_2_1}), power-law density profile with
$q$ = 1 and $q$ = 10 (Eq.~\ref{eq:L_add_pow2_1}), exponential density
profile (Eq.~\ref{eq:L_add_exp}, i.e., $q\to\infty$), and the standard
1D model which ignores photons emitted from $R_{\rm in} < R < R_{\rm
trap}$ (Eq.~\ref{eq:lum_2_2_2}). Triangles, circles, and squares
connected by long-dashed lines indicate the numerical results from
Fig.~7 of Ohsuga~et~al.~(2005) corresponding to metallicity $Z$ = 0,
$1 Z_{\odot}$, and 10 $Z_{\odot}$, respectively. 
%For details seeOhsuga~et~al.~(2005). 
\footnote{The existence of metallicity has two conflicting effects
which influence the disk structure, i) larger radiation coefficient
induced by the sizable amount of metals helps to cool the gas more,
which means that the gas can be easily accreted, and ii) the larger
absorption coefficient by such a metallicity pushes the gas outside,
which means that it is difficult to be accreted. However there is an
uncertainty of metallicity in the system. Therefore, Ohsuga et
al. (2005) studied three cases of the metallicity in a reasonable
range, i.e., $Z / Z_{\odot}$=0, 1, and 10. The variation of the
luminosity which comes from this uncertainty should be treated as an
error in the numerical simulation. For details see
Ohsuga~et~al.~(2005). } The dotted line represents the results of the
hydrodynamical simulations of the slim-disk model carried out by
Watarai et al. (2000), which did not explicitly include the effects
discussed here.  In these calculations, we adopted $R_{\rm in}=3
R_{\rm g}$ and $h$ = 0.4.

%%%%%%%%%%%%%%%%%%%%%%%%%%%%%%%%%%%%%%%%%%%%%%%%%%%%%%%%%%%%%%%%%%%%%%%%
%%%%%%%%%%%%%%%%%%%%%%%%%%%%%%%%%%%%%%%%%%%%%%%%%%%%%%%%%%%%%%%%%%%%%%%

From Fig.~\ref{fig:lm} we see that the luminosity is larger than with
the standard analytical 1D model if we ignore radiation from the
photon-trapped region. The increase is the result of the
multi-dimensional nature of photon diffusion, which allows radiation
to partially escape even from radii $R < R_{\rm trap}$.

For reference, we have also plotted by a short-dashed line the
approximate formula Eq.~(\ref{eq:lum_2_2}) for the disk luminosity in
the 1D model, which includes the photon luminosity emitted from
$R_{\rm in} < R < R_{\rm trap}$ by assuming a simple local Eddington
flux.  Because of the logarithmic functional form of this
approximation, this model resembles the exponential density model in
the 2D treatment.  Correspondingly, we conclude that this model
provides a reasonable 1D approximation.

Ohsuga et al. (2005) pointed out that photons tend to have momenta
pointed inwards because of advection by the accretion flow, which
causes the radiation to spend a longer time in the disk before
escaping relative to the simple diffusion estimate we have used in our
model.  However, we believe this introduces only a minor correction to
our results because we are focusing mainly on photons for which
$t_{\rm diff} < t_{\rm acc}$.

In addition, Ohsuga et al. (2002) obtained outflows driven by the
super Eddington luminosity of their models.  Although our analytical
model does not include outflows, we believe our discussion is still
applicable in a time-averaged mean sense.

%%%%%%%%%%%%%%%%%%%%%%%%%%%%%%%%%%%%%%%%%%%%%%%%%%%%%%%%%%%%%%%%%%%%%%%
\section{Discussion and Conclusion}
%%%%%%%%%%%%%%%%%%%%%%%%%%%%%%%%%%%%%%%%%%%%%%%%%%%%%%%%%%%%%%%%%%%%%%%

In this paper, we have studied photon emission from a supercritical
accretion disk. In the usual one-dimensional treatment of disks, one
integrates all quantities along the vertical ($z$) direction.  This is
reasonable for a thin accretion disk when the accretion rate is well
below the Eddington rate.  However, we show that, when the accretion
rate is large, it is quite important to consider the $z$-dependence of
fluid quantities and to detreat the vertical diffusion of photons
carefully.  If we omit the $z$-dependence of the diffusion process in
the simplest version of the one-dimensional model, we tend to
significantly underestimate the luminosity.  On the other hand, if we
omit the $z$-dependences of the matter density and assume uniform
density, we overestimate the luminosity.  Only when we consider the
$z$-dependence of the matter density and adopt the appropriate density
profile such as a power-law or exponential form as a function of $z$,
as seen in 2D radiation-hydrodynamical simulations, are we able to
reproduce the luminosity found in the numerical simulations.

Using the analytical approach developed here, we should be able to
model how the peak energy of the photon spectrum decreases when the
photons produced near the BH are trapped, as pointed out by Ohsuga et
al. (2003). This will be discussed in a forthcoming paper (Kohri \&
Ohsuga, 2007).

When we simultaneously include radiative, advective, convective, and
neutrino cooling processes in an analytical model of the accretion
disk, the effects discussed here will play a crucial role.  In
neutrino dominated accretion flows (NDAFs), for instance, the neutrino
luminosity and annihilation rate could be modified substantially
relative to the predictions of one-dimensional models.  The dominant
cooling process could be changed, and the entire picture of the
accretion might be dramatically modified, e.g., for the produced
energy through the $\nu$ $\bar{\nu}$ annihilation (Di Matteo, Perna \&
Narayan 2002, Gu, Liu \& Lu 2006, Chen \& Beloborodov), $r$-process
nucleosynthesis (Surman, McLaughlin, Hix 2006), and so on.  We plan to
discuss these effects in forthcoming papers.

Multi-dimensional hydrodynamic simulations of accretion disks
including all of these cooling processes have not been done so far
(except for the 2D numerical simulations of neutrino cooled disks by
Lee \& Ramirez-Ruiz, 2006). To clarify the role of
multi-dimensionality on CDAFs, NDAFs, etc., it would be useful to
compare the analytical estimates obtained in this paper with full
multi-dimensional numerical simulations. It is hoped that these
simulations will be done in the near future.

%%%%%%%%%%%%%%%%%%%%%%%%%%%%%%%%%%%%%%%%%%%%%%%%%%%%%%%%%%%%%%%%%%%%%%%
\section*{Acknowledgements}
%%%%%%%%%%%%%%%%%%%%%%%%%%%%%%%%%%%%%%%%%%%%%%%%%%%%%%%%%%%%%%%%%%%%%%%

K.K. would like to thank A.~E. Broderick, J.~C. McKinney and
S. Mineshige for valuable discussions at an early stage of this work.
This work was supported in part by NASA grant NNG04GL38G, PPARC grant,
PP/D000394/1, EU grant MRTN-CT-2006-035863, the European Union through
the Marie Curie Research and Training Network "UniverseNet"
(MRTN-CT-2006-035863), and Research Grant from Japan Society for the
Promotion of Science (17740111).

%%%%%%%%%%%%%%%%%%%%%%%%%%%%%%%%%%%%%%%%%%%%%%%%%%%%%%%%%%%%%%%%%%%%%%%
\appendix
%%%%%%%%%%%%%%%%%%%%%%%%%%%%%%%%%%%%%%%%%%%%%%%%%%%%%%%%%%%%%%%%%%%%%%%

%%%%%%%%%%%%%%%%%%%%%%%%%%%%%%%%%%%%%%%%%%%%%%%%%%%%%%%%%%%%%%%%%%%%%%%
\section{Analytical estimates of disk-half thickness $H$}
%%%%%%%%%%%%%%%%%%%%%%%%%%%%%%%%%%%%%%%%%%%%%%%%%%%%%%%%%%%%%%%%%%%%%%%
\label{sec:disk-half-thickness}

In this section, we simply try to analytically estimate  the
$R$-dependence on disk-half thickness $H$. The force balance along
$z$-axis is generally expressed by
\begin{eqnarray}
    \label{eq:z_force}
    \frac{\sigma_{T}}{m_{p} c} F = \frac{G M}{R^{2}} \frac{H}{R},
\end{eqnarray}
with a photon flux of $F$.

If $R_{\rm trap} < R$,  then it would be reasonable to assume that the
flux is  equal to the viscous heating rate, $F \approx Q_{\rm vis}$. Then,
we obtain
\begin{eqnarray}
    \label{eq:h_large}
    \left. \frac{H}{R} \right|_{R_{\rm trap} < R}
    &\approx& 
    \frac34  \frac{R_{\rm g}}{ R}  \dot{m}.
\end{eqnarray}
Therefore $H/R$ is approximately proportional to $\propto 1/R$ for
$R_{\rm trap} < R$.

On the other hand, if $R < R_{\rm trap}$, we may assume that the flux
would be approximately the order of the Eddington flux $F \approx
\frac12 F_{\rm E} = \frac12 L_{\rm E}/(4\pi R^{2})$. The factor 1/2 in
front of $F_{\rm E}$ is attached as a matter of convenience for the
consistency. That comes  from the viewpoint of continuities of astrophysical
quantities although physics does not change at all by this artificial
factor. From Eq.~(\ref{eq:z_force}) we find that $H/R$ is constant for
$R < R_{\rm trap}$,
\begin{eqnarray}
    \label{eq:h_small}
    \left. \frac{H}{R} \right|_{R<R_{\rm trap}} \approx \frac12.
\end{eqnarray}

Then we see that $h = H/R$  is a smooth function of  $R$,
\begin{eqnarray}
    \label{eq:h_func}
    h = 
    \left\{
      \begin{array}{cc}
          1/2 & (R \le R_{\rm trap}),\\
          \\
          \frac12 R_{\rm trap}/R &  (R_{\rm trap} < R ),
      \end{array}
    \right.
\end{eqnarray}
if we take $R_{\rm trap} = 3/2 h_{\rm in} \dot{m}
R_{g}$ with $h_{\rm in} = 1/2$.

%%%%%%%%%%%%%%%%%%%%%%%%%%%%%%%%%%%%%%%%%%%%%%%%%%%%%%%%%%%%%%%%%%%%%%%
\section{Derivation of the $z$-dependent/independent diffusion time scale}
%%%%%%%%%%%%%%%%%%%%%%%%%%%%%%%%%%%%%%%%%%%%%%%%%%%%%%%%%%%%%%%%%%%%%%%
\label{sec:deriv_tdiff}

Here we discuss the diffusion time scale  along with $z$-axis in 3D
random walk processes when  the scattering length depends on the
position, i.e., in the case that we consider the  $z$-dependent number
density of electron, the cross section, and so on in the
electron-photon scattering processes.

In general, we can write the summed distance measured along the 3D
random walk process from the stating point $z=z_{0}$ as
\begin{eqnarray}
    \label{eq:dz}
    d(z_{0},z_{\rm max}) =  
    \sum_{i_{z}=1}^{N_{\rm scatt}(z_{0},z_{\rm max})~}
    \int_{z_{i_z-1}}^{z_{i_z}} dz,
\end{eqnarray}
where  $N_{\rm scatt}(z_{0},z_{\rm max})$  is the  number of  the
scattering from $z=z_{0}$  to a  maximum of $z$  ($\equiv z_{\rm
max}$). The  integral interval $[z_{i_z-1}, \  z_{i_z}]$ is determined
by  solving the  following differential  equation for the number of
the scattering,
\begin{eqnarray}
    \label{eq:dNdz}
    \frac{dN}{dz} = \sigma_{\gamma e} n_{e} (z).
\end{eqnarray}
When we solve it,  we may assume that the interval between the
scatterings is determined by the approximate relation,
\begin{eqnarray}
    \label{eq:intdN}
   \left| \int_{z_{i_z-1}}^{z_{i_z}} \sigma_{\gamma e} n_{e} (z) dz
   \right| = \frac13,
\end{eqnarray}
where the  meaning of dividing by three comes from  the contribution
only to $z$-axis in the the 3D  random walk. Of course the right hand
side would not have to be one third if only it were the order of
$\cal{O}$(1). For simplicity,  here we  have just took it one
third. In addition, generally $z_{i_z}$ must not be larger than
$z_{i_z-1}$.

As we will see later, the diffusion tends to be proceed outward with
increasing diffusion length or mean free path. Then the time spent in
inner regions tends to be much longer than that in outer ones. That
means that  phenomena of the diffusion which started from $z=z_{0}$
are mainly determined by the local physics at around $z=z_{0}$. Then
Eq.~(\ref{eq:intdN}) might be rewritten as a definition of the
$z$-dependent mean-free path $\lambda(z)$ through
\begin{eqnarray}
    \label{eq:def_lambda_1}
    \int_{z_{i_z-1}}^{z_{i_z}} \sigma_{\gamma e} n_{e} (z) dz
    \approx \sigma_{\rm T} n_{e} (z_{i_z-1})\int_{z_{i_z-1}}^{z_{i_z}}dz,
%    {\sigma_{\rm T} \ n_{e}(z_{i_z-1})} \  {\lambda(z_{i_z-1}) }  = 1,
\end{eqnarray}
by
\begin{eqnarray}
    \label{eq:def_lambda_0}
    \lambda(z_{i_z-1}) \equiv  \left| \int_{z_{i_z-1}}^{z_{i_z}} dz \right|
    \approx  \frac13 \frac1{  \sigma_{\rm T} \ n_{e}(z_{i_z-1})},
\end{eqnarray}
where we assumed that $\sigma_{\gamma e} =\sigma_{\rm T}$.  Although
there would be a lot of ways to define the $z$-dependent mean-free
path,  the definition in Eq.~(\ref{eq:def_lambda_0}) would be
relatively natural in the current context of the accretion disks
because the phenomena are mainly locally determined.

Then $d(z_{0},z_{\rm max})$ in Eq.~(\ref{eq:dz}) is rewritten as 
\begin{eqnarray}
    \label{eq:dz2}
    d(z_{0},z_{\rm max}) =  
    \sum_{i_{z}=1}^{N_{\rm scatt}(z_{0},z_{\rm max})}
    \ell(z_{i_z-1}),
\end{eqnarray}
with
\begin{eqnarray}
    \label{eq:ell}
    \ell(z_{i_z}) = 
    \left\{
      \begin{array}{cc}
          \lambda (z_{i_z}) & (z_{i_z+1} \ge z_{i_z}),\\
          \\
      -  \lambda (z_{i_z}) &  (z_{i_z+1} < z_{i_z}).
      \end{array}
    \right.
\end{eqnarray}

Next let us consider the averaged value of $d(z)$ after sufficiently a
lot of  tries.
\begin{eqnarray}
    \label{eq:ave_dz}
    \left\langle d(z_{0},z_{\rm max}) \right\rangle 
    = \sum_{i_{z}=1}^{N_{\rm scatt}(z_{0},z_{\rm max})} 
    \left\langle \ell(z_{i_z-1})  \right\rangle,
\end{eqnarray}
where $\langle  X \rangle$  means the average of $X$ after such
trials.  Here we can assume  $\left\langle \ell(z_{i_z-1})
\right\rangle \approx 0$.  That is because we have assumed that the
local physics at around $z=z_{i_z-1}$ determines the phenomena, and
surely then this approximation would not be so bad. Then we see that
the averaged value of $d(z_{0},z_{\rm max})$ vanishes,
\begin{eqnarray}
    \label{eq:ave_dz2}
    \left\langle d(z_{0},z_{\rm max}) \right\rangle 
    \approx 0.
\end{eqnarray}

On the other hand, however, the averaged value of the  square of
$d(z)$ must be finite.
\begin{eqnarray}
    \label{eq:ave_dz_sq}
    \lefteqn{ \left\langle d(z_{0},z_{\rm max})^{2} \right\rangle}
    \nonumber \\
    &=&  \sum_{i_{z}=1}^{N_{\rm scatt}(z_{0},z_{\rm max})} \quad
    \sum_{j_{z}=1}^{N_{\rm scatt}(z_{0},z_{\rm max})}  \left\langle 
    \ell(z_{i_z-1})  \ell(z_{j_z-1})  \right\rangle  \nonumber \\
    &=&  \sum_{i_{z}=j_{z}=1}^{N_{\rm scatt}(z_{0},z_{\rm max})} 
    \left\langle  \ell(z_{i_z-1})^{2} \right\rangle +  
     \sum_{i_{z} \neq j_{z}}  \left\langle 
    \ell(z_{i_z-1})  \ell(z_{j_z-1})  \right\rangle \nonumber  \\
    &\approx&  \sum_{i_{z}=j_{z}=1}^{N_{\rm scatt}(z_{0},z_{\rm max})} 
    \left\langle  \lambda(z_{i_z-1})^{2} \right\rangle,
\end{eqnarray}
where we approximated
\begin{eqnarray}
    \label{eq:second_term}
    \lefteqn{\sum_{i_{z} \neq j_{z}}  \left\langle 
      \ell(z_{i_z-1})  \ell(z_{j_z-1})  \right\rangle }\nonumber \\
    &\approx& 
    \sum_{i_{z} \neq j_{z}} \ell(z_{i_z-1})  \left\langle 
      \ell(z_{j_z-1})  \right\rangle  
    \left( \approx 
    \sum_{i_{z} \neq j_{z}}  \left\langle 
      \ell(z_{i_z-1})    \right\rangle  \ell(z_{j_z-1}) \right) \nonumber \\
    &\approx& 0,
\end{eqnarray}
because the step $i_{z}$ does not correlate with  that of $j_{z}$
among the independent trials for $\left\langle \ell(z_{i_z-1})
\right\rangle = 0$. Now we can approximate $\left\langle
  \lambda(z_{i_z-1})^{2} \right\rangle \approx \lambda(z_{0})^{2}$
because the scatterings at the outer regions  do not frequently occur
and hardly contribute to the summation at all. 

Then from (\ref{eq:ave_dz_sq}), we find that the number of the
scattering of the photon diffused from $z$ to $z_{\rm max}$, i.e.,
$\langle d^{2}(z,z_{\rm max}) \rangle = (\int_{z}^{z_{\rm
max}}dz)^{2}$, is represented by~\footnote{Note that we have removed
the subscript ``0'' in $z$.}
\begin{eqnarray}
    \label{eq:nscatt_z}
    N_{\rm scatt}(z,z_{\rm max}) &\approx& 3 \left[ \sigma_{\rm T} n_{e}(z)
    \right]^{2} \left(\int_{z}^{z_{\rm max}} dz \right)^{2} 
    \nonumber \\
    &\approx& 
    3 \left[ \sigma_{\rm T} \int_{z}^{z_{\rm max}} dz~ n_{e} (z)\right]^{2}
    \nonumber \\
    &=& 3 \tau(z)^{2},
\end{eqnarray}
where we have used the same logic in Eq.~(\ref{eq:def_lambda_1}) to
transform the first line to the second.

Here we get the expression of the $z$-dependent diffusion time scale,
\begin{eqnarray}
    \label{eq:tdiff_z_a}
    t_{\rm diff}(z) &\equiv& N_{\rm scatt} (z,z_{\rm max}) 
    \lambda(z)/c \nonumber \\
    &\approx& 3 \tau(z)^{2} \lambda(z) /c.
\end{eqnarray}
Of course, if we omit the $z$-dependence, we immediately get the
$z$-independent diffusion time scale shown in
Eqs.~(\ref{eq:tdiff_1})~and~(\ref{eq:t_diff_1}).

When we consider the power-law density profiles $\propto (1-z/q
H)^{q-1}$ (or the exponential one as their large-$q$ limit),
surely $t_{\rm diff}(z)$ decreases rapidly as $\propto (1-z/q
H)^{q+1}$ as a function of $z$. This means that the time spent in
the inner regions is  much longer than that in the outer ones, that
validates our assumption in the current context in the accretion disks.

%%%%%%%%%%%%%%%%%%%%%%%%%%%%%%%%%%%%%%%%%%%%%%%%%%%%%%%%%%%%%%%%%%%%%%%
%\vfill\eject
%\begin{appendix}
%%%%%%%%%%%%%%%%%%%%%%%%%%%%%%%%%%%%%%%%%%%%%%%%%%%%%%%%%%%%%%%%%%%%%%%

%%%%%%%%%%%%%%%%%%%%%%%%%%%%%%%%%%%%%%%%%%%%%%%%%%%%%%%%%%%%%%%%%%%%%%%
%\section{Appendix: }
%%%%%%%%%%%%%%%%%%%%%%%%%%%%%%%%%%%%%%%%%%%%%%%%%%%%%%%%%%%%%%%%%%%%%%%
%\label{sec:}

%\end{appendix}

%\clearpage

%

%\appendix
%\section[]{}

%\bsp

\label{lastpage}

\end{document}